\documentclass[aps, twocolumn, prl, showpacs, amsmath, superscriptaddress]{revtex4}
\usepackage{graphicx}
\usepackage{dcolumn}
\usepackage{bm}
\usepackage{amsmath}
\usepackage{subfigure}
\usepackage{amsfonts}
\usepackage{xcolor}
\usepackage{appendix}

\begin{document}

\title{Resonant impurities states-driven topological transition in InN$_{x}$Bi$_{y}$Sb$_{1-x-y}$/InSb quantum well}
\author{Zhi-Gang Song}
\affiliation{School of Electrical and Electronic Engineering, Nanyang Technological University, Singapore 639798, Singapore}
\affiliation{State Key Laboratory of Superlattices and Microstructures, Institute of Semiconductors, Chinese Academy of Sciences, Beijing, China}
\affiliation{Synergetic Innovation Center of Quantum Information and Quantum Physics, University of Science and Technology of China, Hefei, Anhui 230026, China}
\author{Wei-Jun Fan}
\email{ewjfan@ntu.edu.sg}
\affiliation{School of Electrical and Electronic Engineering, Nanyang Technological University, Singapore 639798, Singapore}
\author{Yan-Yang Zhang}
\affiliation{State Key Laboratory of Superlattices and Microstructures, Institute of Semiconductors, Chinese Academy of Sciences, Beijing, China}
\affiliation{Synergetic Innovation Center of Quantum Information and Quantum Physics, University of Science and Technology of China, Hefei, Anhui 230026, China}
\author{Shu-Shen Li}
\affiliation{State Key Laboratory of Superlattices and Microstructures, Institute of Semiconductors, Chinese Academy of Sciences, Beijing, China}
\affiliation{Synergetic Innovation Center of Quantum Information and Quantum Physics, University of Science and Technology of China, Hefei, Anhui 230026, China}

\date{\today }

\begin{abstract}
We demonstrate theoretically that a new system InN$_{x}$Bi$_{y}$Sb$_{1-x-y}$/InSb quantum well(QW)
can realize the topological transition based on the 16-band $k$$\cdot$$p$ model.
Utilizing the
strain introduced by the doped impurities, the band anticrossing induced by the doped nitrogen
and valence band anticrossing induced by the doped bismuth, the band gap of the QW is rapidly
decreased and even becomes negative.
 As a result, the topological transitions arise. Furthermore, the band gap as a function of the concentration of nitrogen and bismuth is calculated, where the negative gap corresponds to the topological phase.
Noting the cancel of strain resulting from the combination of tensile strain introduced by N and
compressive strain introduced by Bi, we can easily tune the ratio of the N and Bi to meet the requirement of strain in the crystal growth procedure.
Our proposal provides a promising
approach for topological insulator in traditional semiconductor system utilizing the semiconductor fabrication technologies.

\end{abstract}
\maketitle


Since the topological insulator has been proposed in HgTe/CdTe quantum well(QW)\cite{Bernevig} and confirmed in transport experiments\cite{Konig},
researchers pay more attention on finding more topological insulator materials in traditional semiconductor systems. The
 InAs/GaSb/AlSb Type-II QW was designed theoretically\cite{Chaoxing Liu} and confirmed
 through the transport experiments\cite{Rui-Rui Du}. Later, more theoretical schemes, including 2-D GaAs with hexagonal symmetry\cite{Sushkov}, InN/GaN QW\cite{M.S.Miao} and Ge/GaAs QW\cite{Dong Zhang}, were proposed.
 Another theoretical scheme was explored in InSb material\cite{Yugui Yao}. It was demonstrated that
normal bulk InSb can be converted into a topologically nontrivial phase with a 2\%-3\% biaxial lattice expansion based on the density functional theory and tight-binding calculation.
A recent proposal designs a p-n junction QW\cite{Haijun Zhang} to inverse the band
utilizing the build in electric field. These two schemes both claim that InSb is indeed
a promising material for topological insulator.

Besides large strains, doping can also be used to engineer the gap of InSb. By varying $x$, the band gap of InN$_{x}$Sb$_{1-x}$ has been tuned to be near zero and even negative experimentally\cite{T.D.Veal,D.H.Zhang,YHZhang}
and explained by the band anticrossing model theoretically\cite{J. Wu,Xiuwen}. Similarly,
the reduction of band gap in InBi$_{y}$Sb$_{1-y}$ is observed on experiments\cite{S.C.Das, D.P.Samajdar, Rajpalke, Y.Z. Gao}
and explained by the valence band anticrossing model\cite{Alberi}.
The N-related resonant states couple with the conduction band(CB) and drives the CB to go down\cite{Xiuwen}.
Similarly, the Bi-related resonant states couple with the valence band(VB) and drive the VB to go up\cite{Alberi}.
Therefore, the incorporation of N and Bi in InSb simultaneously offers a promising proposal to realize a topological insulator.

Here, we propose an innovative semiconductor system InN$_{x}$Bi$_{y}$Sb$_{1-x-y}$/InSb QW grown along the $z$ direction to
 realize the topological transition. Although the band gap of InN$_{x}$Sb$_{1-x}$ can be negative when the concentration of N is
 sufficiently high, it is difficult to grow high quality crystal due to the large strain from such high doping. However, as we will propose in this manuscript, the presence of both N and Bi impurities will make the band inversion (to a negative band gap) easier to happen in two aspects. Firstly, the lifting of VB from Bi doping, and the lowering of CB from N doping, can be simultaneously expected, which greatly increase the possibility of band inversion. Secondly, the threshold doping concentration with small strain in material can be much smaller than those of doping N or Bi alone, which can easily meet the requirement for high quality material growth. Besides, we present the phase diagrams of topological states for QW structures and doping concentrations. Furthermore, Rashba spin splitting(RSS) is extremely considerable when the two kinds impurities are doped. We find that the magnitude of RSS can be the order of 1-2meV when the 0.5mV/nm external electric-field is added. Such a large RSS increases the chance of manipulating the spin\cite{A. Manchon,Dario Bercioux}.

Although density functional theory has been used to study the topological properties in strained bulk InSb\cite{Yugui Yao} and small supercell InSb p-n junction\cite{Haijun Zhang},
here we use 16-band $k$$\cdot$$p$ method that can deal with large supercell realistic quantum well structures.
Following the methodology introduced in Refs.\cite{Alberi, Broderick}, we use a 16-band $k$$\cdot$$p$ Hamiltonian to calculate the QW InN$_{x}$Bi$_{y}$Sb$_{1-x-y}$ band structure by explicitly including the effects of band anticrossing and valence band anticrossing. These two effects representing the interactions between the resonant states and extended states are abstracted from the many-impurity Anderson model\cite{J. Wu}. The 16-band Hamiltonian has the following form:
\begin{equation}\label{eq:h-envelop}
H=\left(
\begin{array}{cccc}
H_{N} & V_{N} & 0 & 0 \\
V_{N}^{*} & H_{CB} & V_{CB} & 0 \\
0 & V_{CB}^{*} & H_{6\times 6} & V_{Bi} \\
0 & 0 & V_{Bi}^{*} & H_{Bi}%
\end{array}%
\right)
\end{equation}

Here, $H_{CB}$ and $H_{6\times 6}$ are the block Hamiltonian of the CB and VB,
$H_{N}$ and $H_{Bi}$ are the Hamiltonian related to the doped impurities N and Bi, respectively. $V_{N}$ is the coupling between
the CB and the resonant N states, and $V_{Bi}$ is the coupling between the VB and the resonant Bi states. The Hamiltonian of
N and Bi impurities states $H_{N}$ and $H_{Bi}$, and the coupling $V_{N}$ and $V_{Bi}$ are constructed for dilute doping case in the coherent-potential approximation\cite{J. Wu}. The
details of the Hamiltonian[Eq. (\ref{eq:h-envelop})] are as follows:
\begin{subequations}
\begin{equation}\label{}
H_{N}=\left(
        \begin{array}{cc}
          E_{N} & 0 \\
          0 & E_{N} \\
        \end{array}
      \right)
\end{equation}
\begin{equation}\label{}
V_{N}=\left(
        \begin{array}{cc}
          \beta_{N}\sqrt{x} & 0 \\
          0 & \beta_{N}\sqrt{x} \\
        \end{array}
      \right)
\end{equation}
\begin{equation}\label{}
H_{Bi}=\left(
         \begin{array}{cccccc}
           E_{Bi}^{HH} & 0 & 0 & 0 & 0 & 0 \\
           0 & E_{Bi}^{LH} & 0 & 0 & 0 & 0 \\
           0 & 0 & E_{Bi}^{LH} & 0 & 0 & 0 \\
           0 & 0 & 0 & E_{Bi}^{LH} & 0 & 0 \\
           0 & 0 & 0 & 0 & E_{Bi}^{SO} & 0 \\
           0 & 0 & 0 & 0 & 0 & E_{Bi}^{SO} \\
         \end{array}
       \right)
\end{equation}
\begin{equation}\label{}
V_{Bi}=\left(
         \begin{array}{cccccc}
           \beta_{Bi}\sqrt{y} & 0 & 0 & 0 & 0 & 0 \\
           0 & \beta_{Bi}\sqrt{y} & 0 & 0 & 0 & 0 \\
           0 & 0 & \beta_{Bi}\sqrt{y} & 0 & 0 & 0 \\
           0 & 0 & 0 &\beta_{Bi}\sqrt{y} & 0 & 0 \\
           0 & 0 & 0 & 0 & \beta_{Bi}\sqrt{y} & 0 \\
           0 & 0 & 0 & 0 & 0 & \beta_{Bi}\sqrt{y} \\
         \end{array}
       \right)
\end{equation}
\end{subequations}
where $E_{N}$ is the energy level of N-related resonant state and $E_{Bi}^{HH}$, $E_{Bi}^{LH}$ and $E_{Bi}^{SO}$
 are the energy level of Bi-related resonant states, respectively.
 The strain-induced shifts to the band edge energies in our QW system
are calculated according to elasticity theory as described in Ref\cite{C. A. Broderick}.  
The parameters related to N and Bi are: $E_{N}=0.65$eV, $\beta_{N}=3.0$eV \cite{Meyer},
$E_{Bi}^{HH}=E_{Bi}^{LH}=-1.2$eV, $\beta_{Bi}=1.35$eV, $E_{Bi}^{SO}=-2.5$eV\cite{Rajpalke}.
All of the InSb-related parameters in $H_{CB}$ and $H_{6\times6}$ can be found in Ref.\cite{Vurgaftman}.


First, we demonstrate the realization of a  topological transition with increasing the width of the QW,
when the concentrations of N and Bi are set as 2.4\% and 3.2\% with nearly 0.48\% tensile strain compared with nearly 1.4\% tensile strain when 6\% N-doped alone\cite{T.D.Veal}. Fig. 1(a)(b)(c)
 show the procedure of  topological transition with increasing QW widths. Since the tensile strain introduced by doped impurities pushes the $\left\vert LH, \uparrow\downarrow \right\rangle$ up to the $\left\vert HH, \uparrow\downarrow \right\rangle$ and closer to the fermi level,
the four bands that participate the topological transition are the $\left\vert CB, \uparrow\downarrow \right\rangle$ and $\left\vert LH, \uparrow\downarrow \right\rangle$. At the beginning with a narrow well in Fig. 1(a), CB is higher than LH.
 With the increasing of the well width, the CB and LH start to touch each other[Fig. 1(b)]. With an even larger width, the CB and LH will be inverted[Fig. 1(c)].
Fig. 1(d) shows the band gap as a function of the width and barrier of the quantum well. This is also a phase diagram of the topological transition, with the positive (negative) gap corresponding to the trivial (topological) insulator.

\begin{figure}
\includegraphics[width=1.1in,trim={0 0 0 -3cm}]{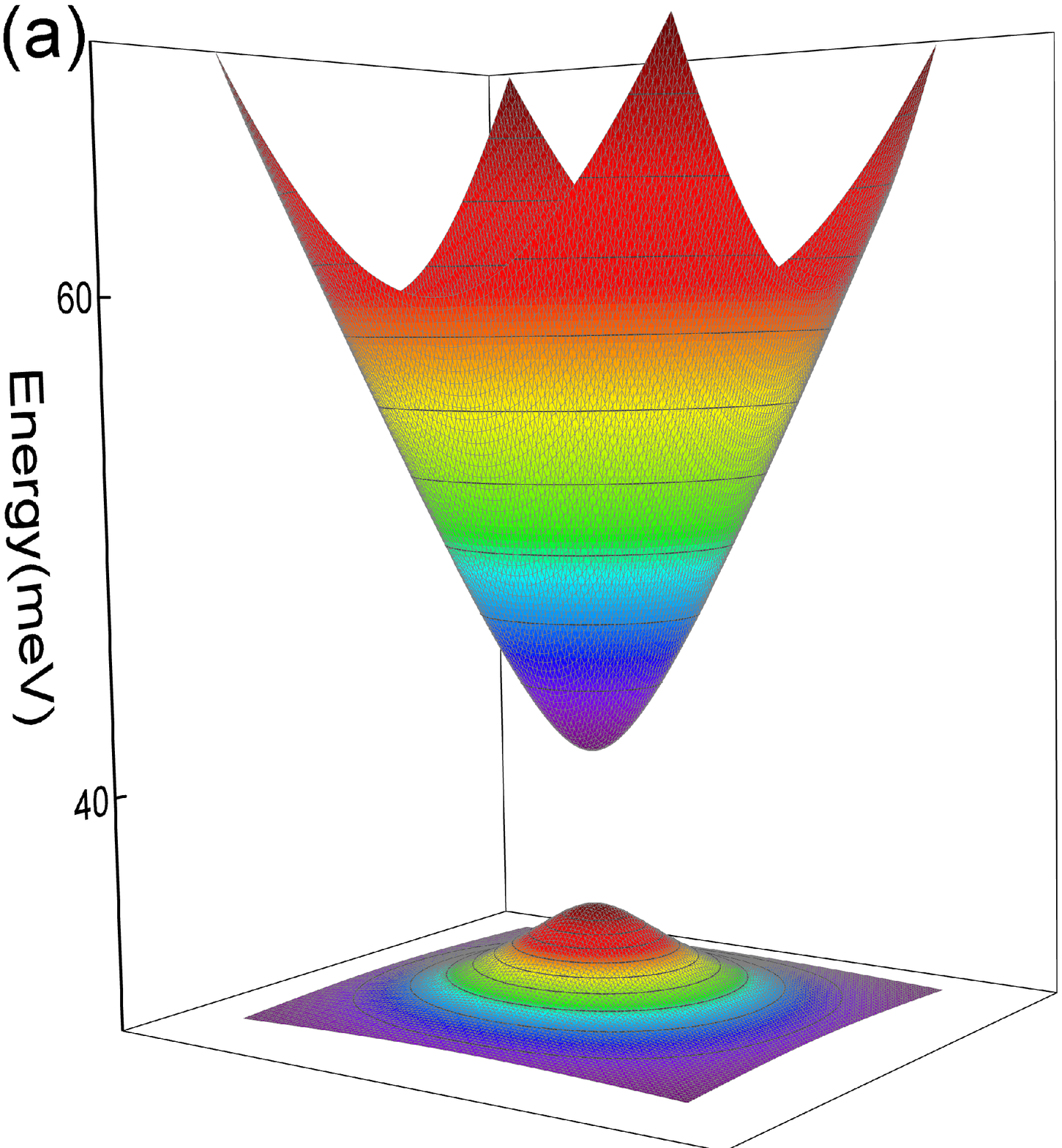}
\includegraphics[width=1.1in,trim={0 0 0 -3cm}]{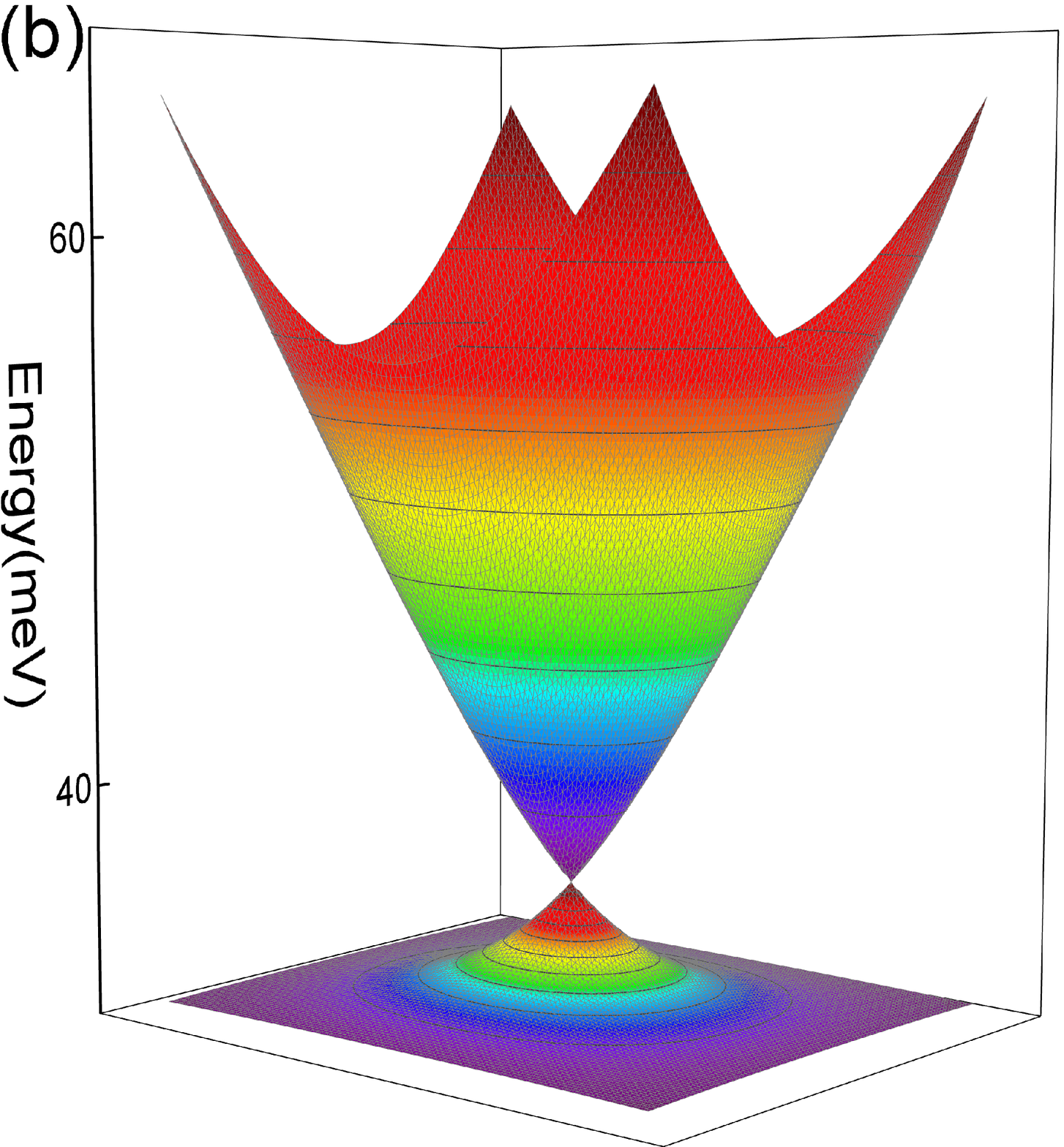}
\includegraphics[width=1.1in,trim={0 0 0 -3cm}]{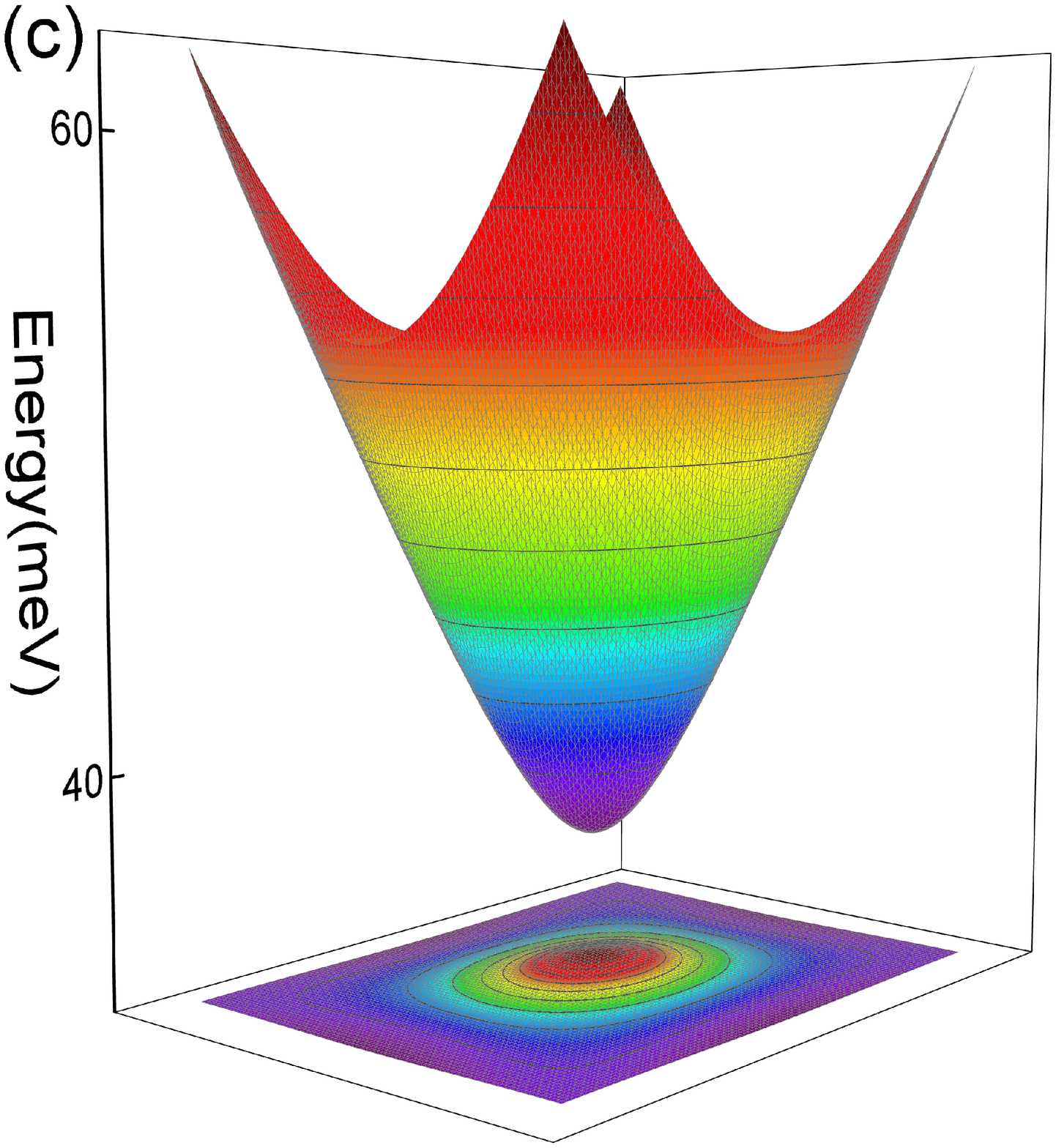}
\includegraphics[width=3.2in]{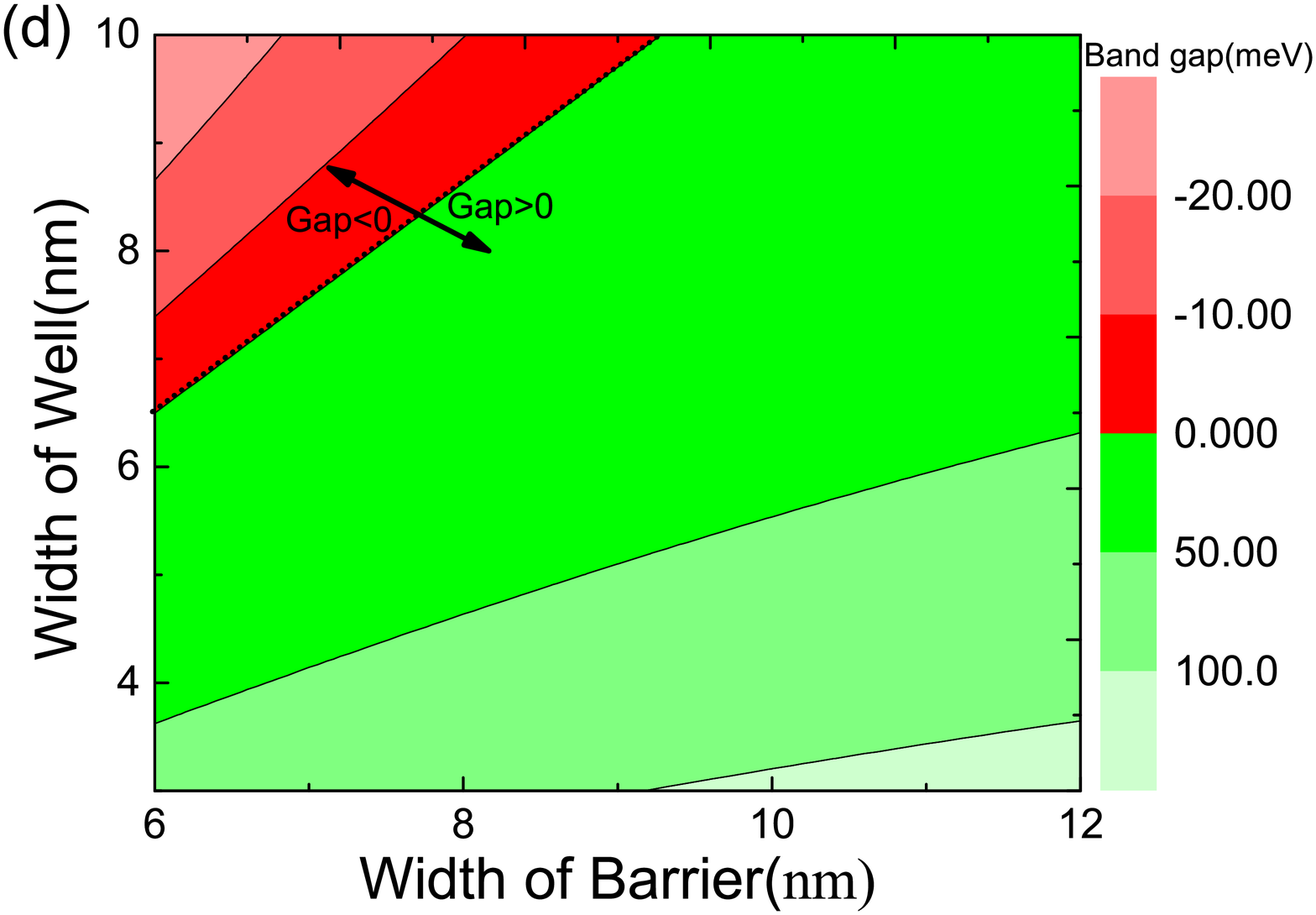}
\caption{
The 2-D band structure around the $\Gamma$ point of the InN$_{x}$Bi$_{y}$Sb$_{1-x-y}$/InSb QW. The width of the well are 7nm, 7.7nm and 8.2nm for (a) (b) and (c), respectively. Obviously, the band gap experience
the process from open and closed and then reopen. This is a typical topological transition. (d)The variation of band gap taken the width of
well and barrier as variables. The whole region are divided into two different phase regions: one is
the positive band gap region and the other is negative. }
\end{figure}

We have seen the band inversion of the QW from the bulk band structures. By using the approach of projecting the original mutilband Hamiltonian [Here is Eq. (\ref{eq:h-envelop})] into
low energy subspace, we will confirm that the phase with inverted bands is indeed a topological phase by the existence of edge states from a quasi-one-dimension geometry. To this end, as in Ref.\cite{D G Rothe} we construct the effective two-dimension(2D) $k$$\cdot$$p$ Hamiltonian by averaging the $z$ component in the 16-band Hamiltonian.
\begin{equation}\label{}
H_{eff}(\textbf{k}_{||})=\langle\psi(z)|H|\psi(z)\rangle
\end{equation}
The Hamiltonian can be divided into
\begin{equation}\label{}
H=H(0,0,-i\frac{\partial}{\partial z})+H^{'}(k_{x},k_{y},-i\frac{\partial}{\partial z})
\end{equation}


We take the $H(0,0,-i\frac{\partial}{\partial z})$ as no-perturbation Hamiltonian and $H^{'}(k_{x},k_{y},-i\frac{\partial}{\partial z})$ as perturbation Hamiltonian.
According to Lowding perturbation method\cite{D G Rothe} we have£º
\begin{widetext}
\begin{equation}\label{}
H_{mm^{'}}^{eff}=E_{m}\delta_{mm^{'}}+H_{mm^{'}}^{'}+\frac{1}{2}\sum_{l}H_{ml}^{'}H_{lm^{'}}^{'}(\frac{1}{E_{m}-E_{l}}+\frac{1}{E_{m^{'}}-E_{l}})
\end{equation}
\end{widetext}
where $m,m^{'}\in{A}$ set and $l\in{B}$ set. Here, the elements in set ${A}$ are the band indexes which we are interested in and the elements in ${B}$ are
beyond our interests. Hence we need to choose appropriate ${A}$ set and ${B}$ set to construct the effective Hamiltonian which can
accurately describe the band structure. As the main participants are the $\left\vert CB, \uparrow\downarrow \right\rangle$ and $\left\vert LH, \uparrow\downarrow \right\rangle$ mentioned above, we ought to include them in A set.
Besides, the HH subbands with the dominant component $\left\vert HH, \uparrow\downarrow \right\rangle$ are about 10meV below the LH subbands.
This is comparable with the bulk gap in energy. Thus, we also need to include them in A set. As a result, our new basis functions, ${A}$ set, are :$\left\vert CB, \uparrow \right\rangle$,
$\left\vert LH, \uparrow \right\rangle$, $\left\vert HH, \uparrow \right\rangle$, $\left\vert CB, \downarrow \right\rangle$,
$\left\vert LH, \downarrow \right\rangle$, $\left\vert HH, \downarrow \right\rangle$. We downfold the 16-band model Hamiltonian to an
2D Hamiltonian expressed in the above six basis. Other subbands from the envelope function expansion coupling with the above six subbands are also essential and should be included in the perturbation for the 2-order
perturbation as they directly influence the coefficients of the $k_{||}^{2}$ terms. Therefore, we include the 10 lowest CB subbands and 20 highest VB subbands into the ${B}$ set.
At last, we successfully construct a 6-band effective Hamiltonian:
\begin{widetext}
\begin{equation}\label{eq:h-eff}
H_{eff}=\left(
\begin{array}{cccccc}
E_{0}+E_{1}k_{||}^{2} & A_{2}k_{-} & A_{1}k_{+} & 0 & 0 & 0 \\
A_{2}^{*}k_{+} & L_{0}+L_{1}k_{||}^{2} & Bk_{+}^{2} & 0 & 0 & 0\\
A_{1}^{*}k_{-} & B^{*}k_{-}^{2} & H_{0}+H_{1}k_{||}^{2} & 0 & 0 & 0 \\
0 & 0 & 0 & E_{0}+E_{1}k_{||}^{2} & -A_{2}k_{+} & -A_{1}k_{-}\\
0 & 0 & 0 & -A_{2}^{*}k_{-} & L_{0}+L_{1}k_{||}^{2} & Bk_{-}^{2}\\
0 & 0 & 0 & -A_{1}^{*}k_{+} & B^{*}k_{+}^{2} & H_{0}+H_{1}k_{||}^{2}%
\end{array}%
\right)
\end{equation}
\end{widetext}
where $k_{||}$ denotes the in-plane momentum, and $k_{\pm}=k_{x}\pm ik_{y}$.
The related parameters are: $E_{0}=0.03165$eV, $L_{0}=0.03784$eV, $H_{0}=0.02017$eV, $E_{1}=0.99574$eV$\cdot$nm$^{2}$, $L_{1}=-0.17255$eV$\cdot$nm$^{2}$,
$H_{1}=-0.09984$eV$\cdot$nm$^{2}$, $A_{1}=-0.4395$eV$\cdot$nm, $A_{2}=-0.2911$eV$\cdot$nm, $B=0.10702$eV$\cdot$nm$^{2}$.

\begin{figure}
\includegraphics[width=3.2in]{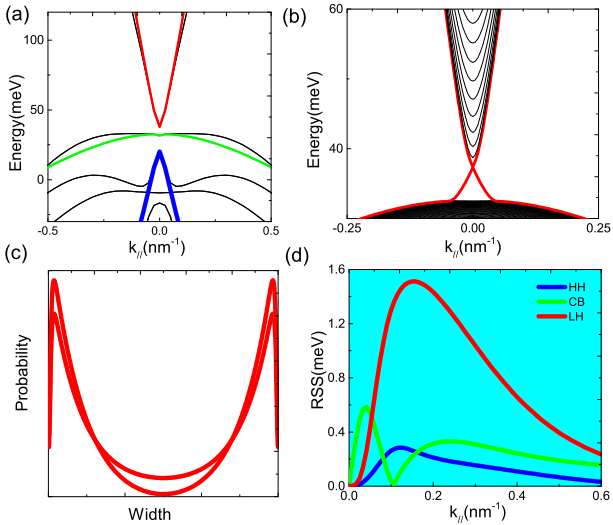}
\caption{(a)The solid black curves describe the band structure of QW InN$_{x}$Bi$_{y}$Sb$_{1-x-y}$/InSb calculated by the 16-band model with envelope functions.
 Besides, the red, green and blue curves(below are the same) which represent the LH, CB and HH are obtained from the 6-band effective Hamiltonian.
 The differences between the 16-band model and 6-band effective model around the $\Gamma$ point is very tiny. This confirms the
rationality of the 6-band effective Hamiltonian.
(b)The band structure of the ribbon based on the 6-band effective Hamiltonian.
The evident red curves are the proofs of the edge states. (c) The distribution of the wave functions of the edge states in the $k_{||}$=0.
The distributions concentrate on the boundary. It also supports the presence of edge states.
(d) The Rashba spin splitting of CB LH and HH. The magnitude of RSS even exceed 1mev for LH. The huge splitting is essential for spintronics.}
\end{figure}

Here, the B-related terms $Bk_{\pm}^{2}$ and $B^{*}k_{\pm}^{2}$ in Eq.(\ref{eq:h-eff}), coupling the HH and LH states, are essential. It is the coupling that introduces the steep profile of LH. Without it,
the band structure calculated by the effective Hamiltonian[Eq. (\ref{eq:h-eff})] could not agree well with the band obtained by $k$$\cdot$$p$ envelop functions[Eq. (\ref{eq:h-envelop})].
To test the validity of the effective Hamiltonian, we compare the band structure calculated by the 16-band model[Eq. (\ref{eq:h-envelop})] and the 6-band effective Hamiltonian[Eq.(\ref{eq:h-eff})].
The tiny differences in Fig. 2(a) indicate that the 6-band effective Hamiltonian can reproduce the profile of band structure near the Fermi level. Based on the effective Hamiltonian (6), we can calculate the band structure of a ribbon geometry, as plotted in Fig. 2(b). The red curves are edge states which connect the CB and VB. In addition, Fig. 2(c) illustrates the localized distributions of their wave functions along the edges. These suggest that they are indeed the topological edge states. Another important term RSS is defined as the splitting of CB or VB when the system is subject to a external electrical field. The RSS of our model is shown in Fig. 2(d), where the external electric-field is $F=0.5$mV/nm. The nonlinear RSS behaviour
agree well with the previous results \cite{Xiuwen}. The magnitude of RSS can be the order of 1-2meV. Such remarkable Rashba spin-orbit coupling offers possibilities of generation, manipulation and detection of spin currents [20, 21].

\begin{figure}
\includegraphics[width=3.2in]{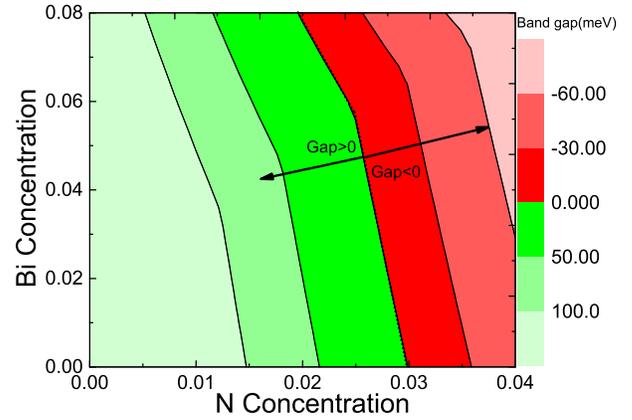}
\caption{The phases of band gap taken the concentration of doped N and Bi as variables. Like Fig1(d), the line between
green and red region divides the whole area into two regions in the view of the sign of band gap. Here,
 we fix the width of well and barrier as 70{\AA} and 82{\AA}. The magnitude of negative band gap can be exceed the
 60meV. This is essential for topological insulator practical implication .}
\end{figure}

All of above have confirmed the topological transition in InN$_{x}$Bi$_{y}$Sb$_{1-x-y}$/InSb QW with varying the well width.
Alternatively, we can also tune the concentration of doping, which is more controllable in the experiments to realize topological transition.
Varying the concentration of doped N and Bi into a reasonable region, the topological transition will arise in our system.
Fig. 3 shows the topological phase diagram about the concentration of impurities. The phase on the right-hand side is topologically nontrivial with a negative gap.
Another feature is that the boundary separating the positive and negative gaps is a broken line. This is different from Fig. 1(d) in which the boundary is a straight line. This is very easy to be understood.
Changing the width of the well and the height of the barrier keeps the strain invariant while different concentrations of impurities lead to tensile or compressive strain. The type of the strain directly determines the highest valence subbands: the tensile(compressive)strain corresponds to $\left\vert LH, \uparrow\downarrow \right\rangle$($\left\vert HH, \uparrow\downarrow \right\rangle$)subbannds as the highest valence subbands. Therefore, the kink of the phase boundary in Fig. 3 corresponds to the alternation between these two cases.

Considering the difficulty of crystal growth for relative large concentration doping, Fig. 3 implies the possibility of growth referred to the
development in dilute nitrogen and bismuth materials. In the view point of semiconductor fabrication technology, although single type impurity doping, i.e., InN$_{x}$Sb$_{1-x}$ or InBi$_{y}$Sb$_{1-y}$ have been successfully grown on InSb substrate, the large strain influences the
 quality of the material. If N and Bi are both incorporation in InSb, the tensile strain introduced by N and compressive strain introduced by Bi can be partly canceled by tuning the ratio of them. This meets the requirement of crystal growth, and the possibility of negative band gap is also increased.
 In addition, if we continue to increase the width of well and decrease barrier relatively, the critical concentration will
decrease more. Thus the possibility of realizing the topological phase is even further increased. Therefore,
our topological insulator scheme in InN$_{x}$Bi$_{y}$Sb$_{1-x-y}$ QW should be experimentally promising.

In summary, we propose a practical scheme to produce topological transition in InN$_{x}$Bi$_{y}$Sb$_{1-x-y}$/InSb QW.
Not only the width of well and barrier, but the concentration of doped N and Bi can be tuned to realize the topological transition.
The effects of the presence of both N and Bi are twofold. First, the strain can be tuned by the ratio of them to adapt to the growth of crystal; Second, the coupling between resonant states and the extent states narrows the band gap until it becomes negative. Here, the resonant states are essential to realize the topological transition. This mechanism is different from other previous schemes. Meanwhile, the large RSS offers the chance of  spin manipulation. In conclusion, our proposal  realizes topological transition utilizing modern semiconductor fabrication in traditional semiconductor material. This is essential for
 topological insulator practical implication.

We are grateful to Xiu-Wen Zhang for value discussion.
W. J. Fan would like to acknowledge the support from MOE Tier 1 funding RG 182/14.
S. S. L would like to acknowledge the support from NSFC Nos. 61427901.


\begin{thebibliography}{*}

\bibitem{Bernevig} B. A. Bernevig, T. L. Hughes, and S. C. Zhang, Science \textbf{314}, 1757 (2006).

\bibitem{Konig} M. Konig, S. Wiedmann, C. Brune, A. Roth, H. Buhmann, L. W. Molenkamp, X.-L. Qi, and S.-C. Zhang, Science \textbf{318}, 766 (2007).

\bibitem{Chaoxing Liu}Chaoxing Liu, T. L. Hughes, X.-L. Qi, Kang Wang and S, C, Zhang, Phys. Rev. Lett. \textbf{100}, 236601(2008).

\bibitem{Rui-Rui Du} Ivan Knez and Rui-Rui Du, Phys. Rev. Lett. \textbf{107}, 136603 (2011).

\bibitem{Sushkov}O. P. Sushkov and A. H. Castro Neto, Phys. Rev. Lett. \textbf{110}, 186601 (2013).

\bibitem{M.S.Miao} M. S. Miao, Q. Yan, C. G. Van de Walle, W. K. Lou, L. L. Li and K. Chang Phys. Rev. Lett. \textbf{109}, 186803 (2012).

\bibitem{Dong Zhang} Dong Zhang, Wenkai Lou, Maosheng Miao, S. C. Zhang and K. Chang Phys. Rev. Lett. \textbf{111}, 156402 (2013).

\bibitem{Yugui Yao} Wanxiang Feng, Wenguang Zhu, H. H. Weitering, G. M. Stocks, Yugui Yao and Di Xiao Phys. Rev. B \textbf{85}, 195114 (2012).

\bibitem{Haijun Zhang}Haijun Zhang, Yong Xu, Jing Wang, Kai Chang, and Shou-Cheng Zhang, Phys. Rev. Lett. \textbf{112}, 216803 (2014).

\bibitem{T.D.Veal}T. D. Veal, I. Mahboob, and C. F. McConville, Phys. Rev. Lett. \textbf{92}, 136801 (2004).

\bibitem{D.H.Zhang} Y. J. Jin, X. H. Tang, J. H. Teng and D. H. Zhang£¬J. Cryst. Growth, \textbf{416}, 12-16 (2015).

\bibitem{YHZhang}Y. H. Zhang, P. P. Chen, H. Yin, T. X. Li and W. Lu, J. Phys. D: Appl. Phys. \textbf{43}, 305405 (2010).

\bibitem{J. Wu} J. Wu, W. Walukiewicz and E. E. Haller, Phys. Rev. B \textbf{65}, 233210 (2002).

\bibitem{Xiuwen} X. W. Zhang, W. J. Fan, S. S. Li and J. B. Xia, Phys. Rev. B \textbf{75}, 205331 (2007).

\bibitem{S.C.Das}S. C. Das, T. D. Das and S. Dhar Infrared Phys. Technol. \textbf{55}, 306-308 (2012).

\bibitem{D.P.Samajdar}D.P.Samajdar and S. Dhar Phys. B. \textbf{484}, 27-30 (2016).

\bibitem{Rajpalke}M. K. Rajpalke, W. M. Linhart, K. M. Yu, M. Birkett, J. Alaria, J. J. Bomphrey, S. Sallis, L. F. J. Piper, T.S. Jones, M. J. Ashwin, and T. D. Veal, Appl. Phys. Lett. \textbf{105}, 212101 (2014).

\bibitem{Y.Z. Gao}Y. Z. Gao, T. Yamaguchi, Cryst. Res. Technol. \textbf{34}, 285 (1999).

\bibitem{Alberi}K. Alberi, J. Wu, W. Walukiewicz, K. M. Yu, O. D. Dubon, S. P. Watkins, C. X. Wang, X. Liu, Y.-J. Cho, and J. Furdyna, Phys. Rev. B, \textbf{75}, 045203 (2007).

\bibitem{A. Manchon}A. Manchon, H. C. Koo, J. Nitta, S. M. Frolov and R. A. Duine, Nature Mater. 14, 871¨C882 (2015).

\bibitem{Dario Bercioux}Dario Bercioux and Procolo Lucignano, Rep. Prog. Phys. 78 106001 (2015).

\bibitem {Vurgaftman}I. Vurgaftman, J. R. Meyer and L. R. Ram-Mohan, J. Appl. Phys. \textbf{89}, 5815 (2001).

\bibitem {Meyer}I. Vurgaftman, J. R. Meyer, Journal of Applied Physics \textbf(94), 3675 (2003).

\bibitem{Broderick}C. A. Broderick, M. Usman and E. P. O'Reilly, Semicond. Sci. Technol. \textbf{28}, 125025 (2013).

\bibitem{C. A. Broderick}C. A. Broderick, P. E. Harnedy, P. Ludewig, Z. L. Bushell, K. Volz, R. J. Manning, and E. P. O'Reilly, Semicond. Sci. Technol. \textbf(30), 094009 (2015).

\bibitem{D G Rothe}D. G. Rothe, R. W. Reinthaler, C-X Liu, L. W. Molenkamp, S-C. Zhang and E. M. Hankiewicz, New J. Phys. \textbf{12}, 065012 (2010).












\end{thebibliography}
\end{document}